\documentclass[preprint,review,12pt]{elsarticle}

\usepackage{amssymb}

\usepackage{amsmath}
\usepackage{url}
\usepackage{multirow}
\usepackage{booktabs}

\journal{Nuclear Physics B}

\begin{document}

\begin{frontmatter}



\title{ShellForge: Adversarial Co-Evolution of Webshell Generation and Multi-View Detection for Robust Webshell Defense}


\author[1]{Yizhong Ding}
\ead{chaoqunding5@gmail.com}



\address[1]{Beijing Electronic Science and Technology Institute}

\cortext[cor1]{Corresponding author}


\begin{abstract}
Webshells remain a primary foothold for attackers to compromise servers, particularly within PHP ecosystems. However, existing detection mechanisms often struggle to keep pace with rapid variant evolution and sophisticated obfuscation techniques that camouflage malicious intent. Furthermore, many current defenses suffer from high false-alarm rates when encountering benign administrative scripts that employ heavy obfuscation for intellectual property protection.
To address these challenges, we present \textbf{ShellForge}, an adversarial co-evolution framework that couples automated webshell generation with multi-view detection to continuously harden defensive boundaries. The framework operates through an iterative co-training loop where a generator and a detector mutually reinforce each other via the exchange of hard samples. The generator is optimized through supervised fine-tuning and preference-based reinforcement learning to synthesize functional, highly evasive variants. Simultaneously, we develop a multi-view fusion detector that integrates semantic features from long-string compression, structural features from pruned abstract syntax trees, and global statistical indicators such as Shannon entropy. To minimize false positives, ShellForge utilizes a LLM-based transformation to create de-malicious samples—scripts that retain complex obfuscation patterns but lack harmful payloads—serving as high-quality hard negatives during training.
Evaluations on the public FWOID benchmark demonstrate that ShellForge significantly enhances defensive robustness. Upon convergence, the detector maintains a 0.981 F1-score while the generator achieves a 0.939 evasion rate against commercial engines on VirusTotal.

\end{abstract}




\end{frontmatter}

\section{Introduction}
\begin{figure}[t]
\centering
\includegraphics[width=0.95\linewidth]{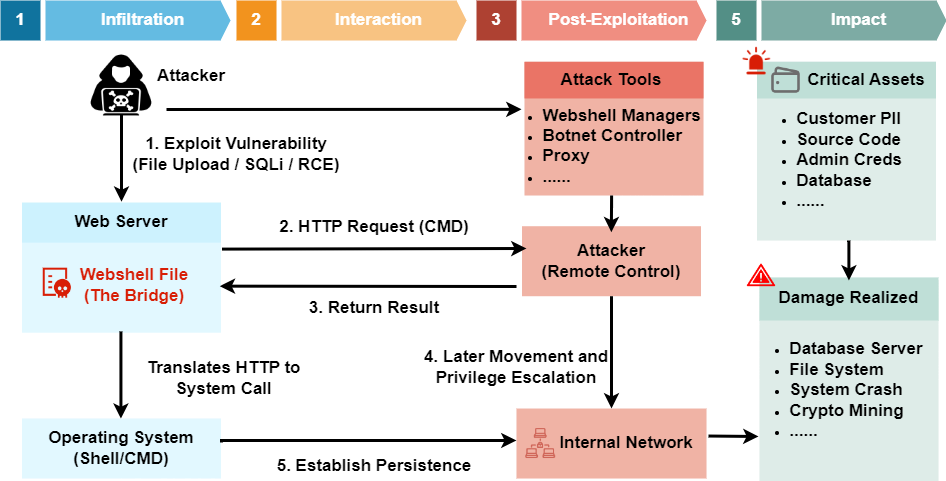}
\caption{An illustration of typical webshell threats and attack workflow in PHP environments.}
\label{fig:webshell_intro}
\end{figure}

Over the past decade, webshells have persisted as a reliable, low-cost foothold for adversaries to compromise and maintain control over internet-facing servers, particularly within PHP-centric environments \cite{ma2024research,yang2018webshell}. Recent incident-response data underscores that webshell implantation is no longer an outlier but a primary post-exploitation primitive for establishing long-term persistence; for instance, Sophos reported that webshells served as a persistence mechanism in 19.74\% of all investigated cases \cite{sophos2023aar-businessleaders}. Government advisories further emphasize the agility of webshell toolchains in bypassing modern mitigations—CISA recently documented threat actors deploying multiple Ivanti-targeting variants that successfully evaded standard integrity-checking mechanisms \cite{cisa2024aa24-060b}. This rapid adaptation is fueled by the integration of multi-layer obfuscation, dynamic invocation, and accelerated variant iteration, which collectively cause traditional signature- and rule-based defenses to lag behind the evolving threat landscape \cite{nsa_asd2020_webshell_csi}. As illustrated in Figure~\ref{fig:webshell_intro}, these tools exploit legitimate web functionalities to provide unauthorized access, serving as a critical bridge for executing system-level actions and compromising server integrity.

Despite sustained progress in PHP webshell defense, a significant gap persists between high laboratory benchmark scores and real-world robustness against rapid obfuscation drift. While existing detection studies leverage statistical cues, behavioral patterns, structure-aware program representations, and Large Language Model (LLM) analysis \cite{li2024identity,li2025tuni,cheng2025gibberish,cheng2024gibberish}, their effectiveness remains brittle when attackers alter surface forms while preserving malicious functionality \cite{xie2024php,wang2024webshell}. Community surveys further highlight a disproportionate bias toward PHP and a lack of standardized, comprehensive benchmarks for cross-validation \cite{hannousse2021handling}. Furthermore, modern webshells exhibit long-context characteristics—such as multi-layer encodings, extensive string literals, and inert code padding—that push detectors into truncation regimes, necessitating sliding-window or incremental defensive approaches \cite{dong2024ast}. On the generation side, obfuscation frameworks relying on preset transformations are often restricted by rigid templates \cite{hu2022generating}. While prompt-based LLM generation offers flexibility, it can be redundant \cite{peiselfprompt,pei2025selfprompt,cao2025agr,cheng2025ecoalign,cao2024agr,cheng2025talk}, explores only a narrow portion of the potential obfuscation space, and is frequently curtailed by provider safety policies. In contrast, reward-driven trainable generators can optimize for diversity and evasiveness but are typically treated as standalone red-teaming tools \cite{beutel2024diverse}. Consequently, the field lacks a principled, webshell-specific environment where generation and detection mutually inform one another to proactively address emerging obfuscation families.

This gap becomes more pronounced as attacker and defender dynamics increasingly resemble an arms race \cite{song2023jshelldetector}: defenders largely rely on models trained on static corpora and evaluated under fixed data splits, while adversaries iteratively mutate payload structures and surface forms to evade both signatures and learned heuristics. Recent studies indicate that even powerful foundation models require task-specific feature extraction and behavioral priors to remain effective, as performance often hinges on the amount of context preserved around risky behaviors. Meanwhile, although trainable generators can be optimized with reward signals to explore vast evasion spaces, the absence of an explicit feedback loop to reveal detector blind spots causes them to risk optimizing against proxy objectives or drifting toward non-representative artifacts. Current adversarial training attempts in the webshell domain remain limited by handcrafted obfuscation operators and fail to achieve sustained mutual adaptation between generation and detection. These tensions highlight an urgent need for webshell-oriented adversarial frameworks that couple generator and detector updates through hard-sample feedback. Such a framework must preserve obfuscation realism while controlling malicious behaviors for safe training and address long-context obfuscation without sacrificing the behavioral cues essential for reliable detection.

Another practical gap is the high false-positive rate of obfuscation-driven detectors in real-world deployments. Prior research has identified an ``obfuscation bias'' in static webshell detection, where treating obfuscation as a primary indicator of maliciousness inadvertently increases the likelihood of misidentifying benign, obfuscated files as threats \cite{lee2024obfuscated}. This issue is a tangible concern: in production environments, benign code is often heavily obfuscated for anti-tampering, license protection, or script bundling. For instance, commercial security plugins frequently utilize hardened code that closely resembles suspicious patterns \cite{sridhar2022backup}. Consequently, detectors that over-rely on surface artifacts—such as long encoded strings or the presence of risky APIs like eval, assert, or base64\_decode—frequently trigger alerts on benign scripts or those with minimal executable behavior. Building a robust defense therefore requires more than just catching evasive webshells; it necessitates learning to localize the specific behaviors that cause harm rather than simply memorizing the surrounding obfuscation scaffolding.

To address these challenges, we propose ShellForge, an adversarial co-evolution framework that tightly couples webshell generation and multi-view detection within a continuous training loop. The process begins with a trainable webshell generator initialized on the FWOID dataset and refined through supervised fine-tuning \cite{wang2025introspective,cheng2025hair,cheng2025inverse,cheng2024rlrf}, followed by preference-based reinforcement learning using a chosen-rejected paradigm \cite{cheng2404reinforcement,cheng2024reinforcement,cheng2025llm,cheng2024deceiving,cheng2025ecoalign}. This approach enables the synthesis of complex malicious code that surpasses the limitations of simple prompt engineering. On the defensive side, we develop a multi-view fusion detector that integrates semantic representations from a CodeBERT-based encoder—incorporating long-string semantic compression to prevent context truncation—with structural signals from pruned abstract syntax trees parsed by tree-sitter and global statistical indicators such as Shannon entropy \cite{shannon1948mathematical} and risky-function usage rates.
The core of ShellForge is an iterative adversarial co-training procedure where generated samples undergo a transformation step using a large language model to remove malicious behaviors while preserving their complex obfuscation patterns. These transformed samples serve as high-quality negatives to help the detector distinguish between harmful intent and benign obfuscation. During each round, samples successfully caught by the detector are fed back to the generator as signals for improvement, while samples that evade detection are injected into the detector's training set as evasive cases. This exchange enables the systematic discovery of defensive blind spots and the hardening of the model against evolving threats. Experimental evaluations demonstrate that this co-evolutionary approach allows the detector to maintain a 0.981 F1-score while the generator achieves a 0.939 evasion rate against commercial engines on VirusTotal.

Our contributions are as follows:
\begin{itemize}
\item We propose ShellForge, the first adversarial framework for Webshell detection that co-trains a generator and detector through iterative updates, enabling the continuous discovery of evasive threats and mutual model hardening.
\item We design a multi-view fusion detector that integrates semantic compression, structural AST signals, and global statistics to effectively overcome sophisticated obfuscation and context truncation issues. 
\item Extensive experiments demonstrate that our safety-preserving de-malicious transformation effectively reduces false alarms on benign-but-obfuscated code and hardens the system against significant distribution shifts. 
\end{itemize}


\section{Related Work}

\subsection{Webshell Detection}

Webshell detection has been studied from multiple perspectives, largely differing in the program representation and the extent to which the method can tolerate obfuscation. Early and lightweight systems commonly relied on lexical signatures and engineered statistics (e.g., entropy, string features, and keyword patterns), often combined with conventional classifiers to achieve strong accuracy on in-distribution data \cite{fang2018detecting,kang2020rf}. To reduce sensitivity to surface-level token changes, many works transform PHP scripts into intermediate representations, most notably opcode sequences. Opcode-based pipelines standardize away part of the syntactic variability and can improve robustness to common encoding and string-manipulation tricks, while still enabling efficient learning with shallow or deep models \cite{fang2018detecting,kang2020rf}.

Structure-aware detection further leverages the syntactic and semantic organization of code beyond flat sequences. AST-based methods serialize abstract syntax trees and apply classic vectorization such as $n$-gram and TF--IDF or neural encoders to capture structural patterns that remain relatively stable under superficial rewriting \cite{gogoi2022php}. Graph-based approaches push this direction by modeling interprocedural control and data dependencies; by extracting and pruning interprocedural control flow graphs (ICFGs) and weighting risky statements, GNN-based detectors can better focus on execution-relevant paths under obfuscation \cite{feng2024glareshell}. However, structure extraction and graph learning also introduce non-trivial preprocessing cost and can be sensitive to parsing failures under adversarially crafted inputs.

More recently, large language models have been explored as general-purpose code analyzers for webshell detection, but their effectiveness depends strongly on how code context is selected and normalized. Behavior-aware strategies that explicitly extract code regions around risky behaviors and reweight function-level evidence can substantially improve LLM-based detection over naive prompting or direct full-file input \cite{han2025can}. However, LLM-centric pipelines can incur substantial computational overhead and latency due to long-context inference and multi-stage prompting, which limits their practicality for large-scale, time-sensitive scanning. A practical constraint in this domain is long-context handling: real-world webshells often include multi-layer encodings, large strings, and inert code padding that can exceed model input limits. Sliding-window attention and incremental learning strategies have been proposed to mitigate truncation and adapt to evolving variants without full retraining \cite{wang2025incremental}. These results suggest that robust detection benefits from combining semantic understanding, structural cues, and lightweight global signals under strict context budgets.

\subsection{Webshell Generation}

Webshell generation research has been motivated by the need for realistic evaluation data under obfuscation drift, as public corpora are often incomplete, noisy, or weakly labeled \cite{hannousse2021handling}. Existing generation techniques range from template-driven mutation to learning-based synthesis. Template and operator-composition approaches apply hand-designed obfuscation primitives (e.g., encoding, string rewriting, dead-code insertion) to existing samples to produce variants, but they can remain constrained by the expressiveness of predefined operators and may struggle to create qualitatively new evasion patterns \cite{pang2023cwsogg}. Prompt-based LLM generation offers a more flexible interface for synthesizing diverse payloads, yet it can be redundant, cover only a narrow portion of the obfuscation space, and is increasingly constrained in practice by strengthened safety alignment and policy enforcement that lead models to refuse or sanitize requests for malicious code. 

Trainable open-source LLMs provide an alternative that supports domain adaptation and controllable exploration through fine-tuning and reinforcement learning. Reward-driven training schemes that combine SFT with PPO-style preference optimization have shown that generators can be steered toward higher stealth and diversity while remaining executable, enabling systematic stress testing beyond brittle prompt engineering \cite{ding2025reward}. In this work, the generator follows this trainable paradigm and is used primarily as a component for improving defenses rather than as a standalone offensive capability.

\subsection{Adversarial Co-Training and Automated Red-Teaming}

Adversarial learning has a long history in security \cite{duan2025oyster,zhao2025strata,cheng2025pbi,chengpbi}, typically framing the interaction between attacker and defender as a data-generation process that exposes model blind spots. In the webshell domain, CWSOGG \cite{pang2023cwsogg} is a representative attempt that uses a GA-based obfuscator as a generator and trains detectors against generated variants in a GAN-like fashion. While such approaches can improve robustness to the particular family of handcrafted transformations, the resulting adversarial pressure is bounded by the operator set and may not reflect the rapidly evolving obfuscation strategies seen in real intrusions.

Beyond webshells, automated red-teaming for malware and security classifiers has studied adversarial example generation and min--max training, including learning-based generators that search for evasive variants under black-box or gray-box constraints \cite{hu2022generating,demetrio2019explaining}. In parallel, LLM-based red-teaming \cite{cheng2025speaker,cheng2025usmid,cheng2024unimodal} has begun to shift from static prompt engineering toward reward-driven optimization that explores a broader attack space with explicit feedback signals \cite{ding2025reward}. These trends motivate adversarial co-training designs that treat generation and detection as mutually reinforcing components, while also requiring domain-specific safeguards that control malicious behaviors during training and evaluation.

\begin{figure}[t]
\centering
\includegraphics[width=1\linewidth]{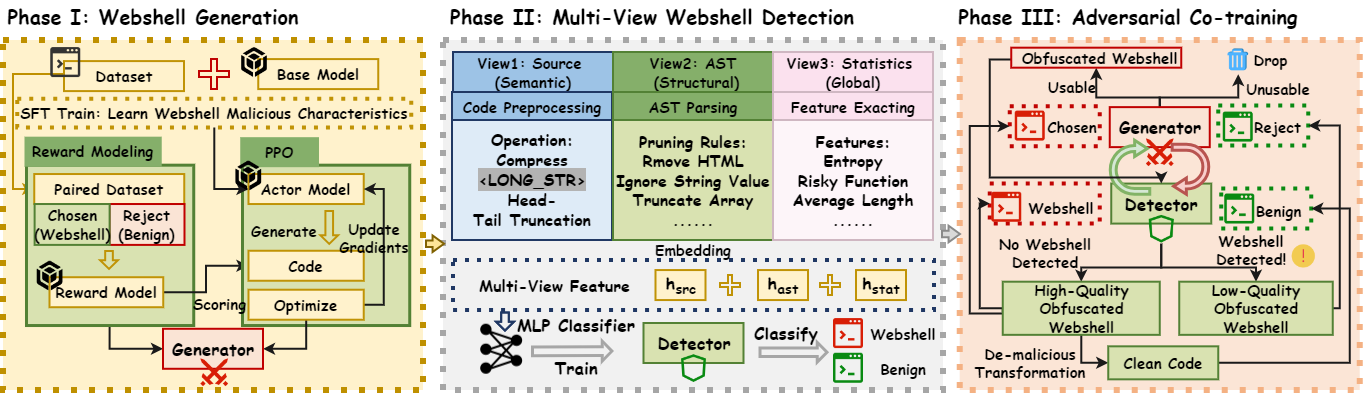}
\caption{Pipeline of the proposed ShellForge framework. We first initialize a trainable webshell generator and a multi-view detector on FWOID. We then perform iterative adversarial co-training driven by detector feedback, where evasive webshells and their de-malicious benign samples are exchanged as hard samples to update the detector and provide chosen/rejected signals for the next generator update. The process continues until both detector robustness and generator evasiveness plateau.}
\label{fig:overview}
\end{figure}

\section{Methodology}
\label{sec:methodology}

This section presents the proposed ShellForge methodology, covering the detector architecture and the adversarial co-training procedure that jointly improve robustness against evolving, heavily obfuscated PHP webshells. The overall pipeline initializes a trainable webshell generator, trains a multi-view detector that fuses semantic, structural, and statistical evidence, and runs an iterative adversarial co-training procedure that performs safe variant transformation and hard-sample exchange between the generator and detector. Figure~\ref{fig:overview} illustrates the overall pipeline of ShellForge.

\subsection{Problem Setup and Notation}

Let a PHP script be denoted as raw code $\mathbf{c}$. We represent $\mathbf{c}$ using three complementary views:
\begin{align}
    \mathbf{x} &= \mathrm{Tok}\big(\mathrm{Compress}(\mathrm{Pre}(\mathbf{c}))\big), \\
    \mathbf{a} &= \mathrm{ASTSeq}\big(\mathbf{c}\big), \\
    \mathbf{s} &= \mathrm{Stats}\big(\mathbf{c}\big),
\end{align}
where $\mathbf{x}$ is the tokenized source-view sequence, $\mathbf{a}$ is the linearized AST-view sequence, and $\mathbf{s}$ is a vector of global statistical indicators. The detector predicts a binary label $y\in\{0,1\}$, where $y{=}1$ indicates a webshell and $y{=}0$ indicates benign. Given a training set $\mathcal{D}=\{(\mathbf{c}_i,y_i)\}_{i=1}^{N}$, we learn a detector $D_{\psi}$ and, for adversarial co-training, maintain a generator $G_{\phi}$ that produces candidate webshell variants.

\subsection{Webshell Generator Initialization}

ShellForge uses a trainable generator $G_{\phi}$ as a controllable source of diverse, functional webshell variants. The generator training pipeline follows our prior work and is only summarized here. We initialize $G_{\phi}$ via SFT on webshell samples to acquire basic PHP-webshell patterning, then improve evasiveness via reinforcement learning with a PPO-style update \cite{schulman2017proximal}. The preference signal is implemented by treating webshell code as \textit{chosen} and benign code as \textit{rejected}, encouraging generations that preserve webshell-like behavior while increasing stealth. 
In this paper, the generator primarily serves the new co-training framework rather than being a standalone contribution.

\subsection{Multi-View Fusion Webshell Detector}

\begin{figure}[t]
\centering
\includegraphics[width=0.95\linewidth]{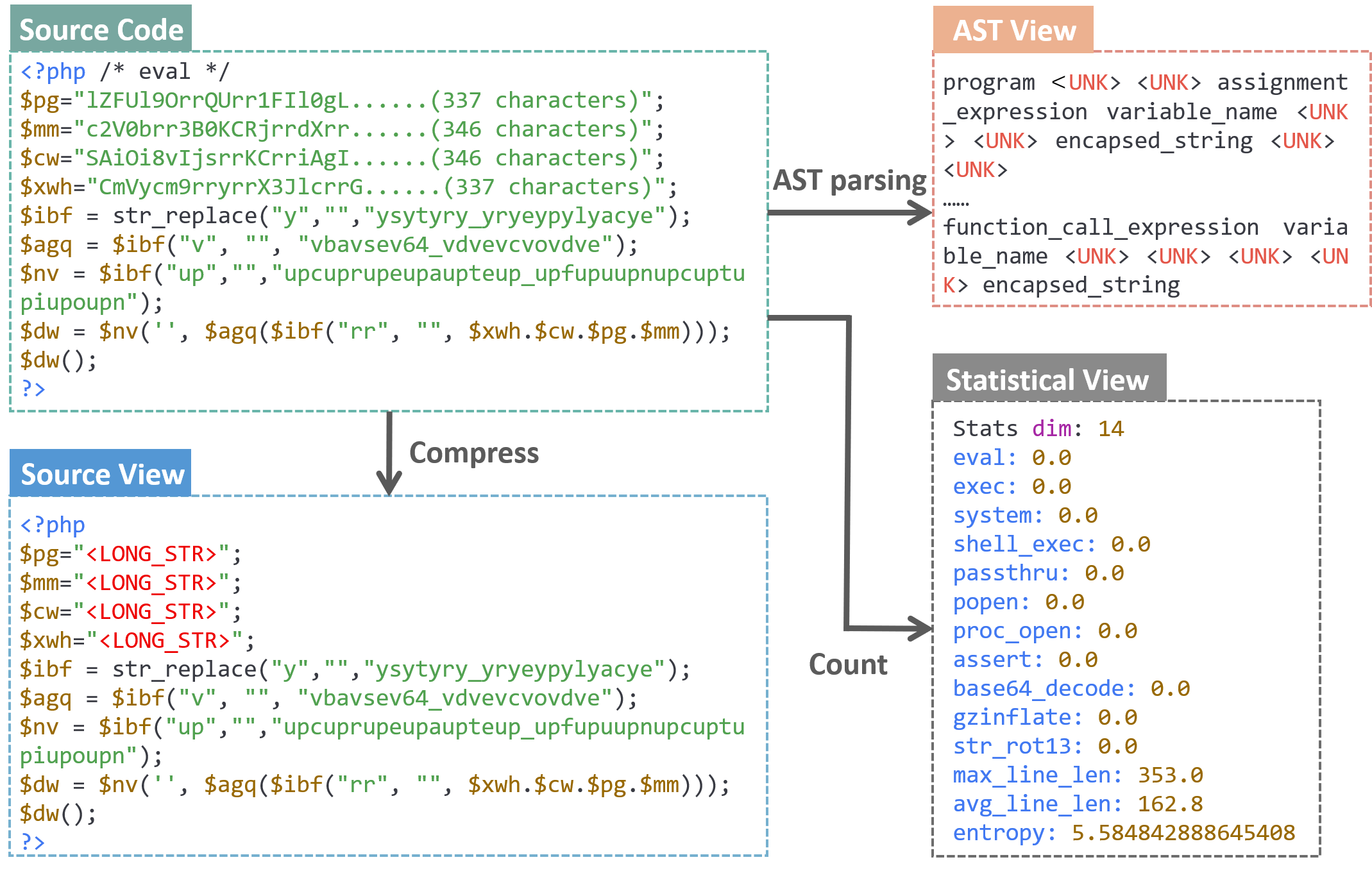}
\caption{Data processing pipeline for the three-view representation. Starting from raw PHP code, we construct the source view via comment removal and long-string semantic compression, the AST view via Tree-sitter parsing with structure pruning, and the statistical view via global indicators such as entropy and risky-function usage.}
\label{fig:data_processing}
\end{figure}

We design a detector that fuses three views to improve robustness under obfuscation, long-context truncation, and structure-preserving transformations. The detector includes a semantic branch over tokenized source code, a structural branch over pruned AST sequences, and a statistical branch over global indicators. The resulting features are concatenated and classified with a multilayer perceptron. Figure~\ref{fig:data_processing} illustrates how a raw PHP script is transformed into the three views.

\subsubsection{Source View: Semantic Compression and Code Encoder}

The source view targets semantic cues while mitigating spurious variance introduced by extreme string obfuscation and excessive length.First, raw code is preprocessed by removing comments and normalizing whitespace. Subsequently, a semantic compression operator $\mathrm{Compress}(\cdot)$ is applied to replace overly long string literals with a placeholder token:
\begin{align}
    \mathrm{Compress}(\mathrm{Pre}(\mathbf{c})) = \mathbf{c}' \quad \text{s.t. long string literals} \Rightarrow \texttt{<LONG\_STR>}.
\end{align}
In our implementation, a string literal is replaced when its length exceeds a threshold $\tau$. Following compression, the code is tokenized using a code-aware encoder \cite{feng2020codebert}. Because webshell files can exceed the input limit, we apply a head--tail truncation strategy: for a token sequence $\{t_j\}_{j=1}^{n}$ and a budget $m$, if $n>m$ we keep the first $\lfloor m/2\rfloor$ tokens and the last $\lceil m/2\rceil$ tokens to preserve both initialization logic and tail-end payload code. The token sequence is then embedded by the encoder and passed to a lightweight convolutional sequence encoder inspired by TextCNN \cite{kim2014convolutional} to obtain a fixed-dimensional semantic feature vector $\mathbf{h}_{\text{src}}$.

\subsubsection{AST View: Tree-sitter Parsing and Structure Pruning}

The AST view targets syntactic structure that is more stable than surface tokens under many obfuscation operations. We parse PHP code using Tree-sitter with a PHP grammar \cite{brunsfeld2018treesitter,treesitterphp}. The parse tree is then linearized with a depth-first traversal that records node \emph{types} only, yielding a sequence $\mathbf{a}=\{a_j\}_{j=1}^{L_a}$.

To reduce noise and improve robustness, we apply three pruning rules during linearization. First, we remove irrelevant HTML or inline text nodes. Second, for string and interpolated-string nodes, we keep the node type but do not traverse their children, preventing large encoded payloads from dominating the structure view. Third, we truncate extremely long array literals by limiting the number of array elements retained and inserting a special truncation marker. We keep at most 10 elements by default. The resulting AST-type sequence is mapped to embeddings and encoded by a lightweight CNN-based sequence encoder to produce a structural feature vector $\mathbf{h}_{\text{ast}}$.

\subsubsection{Statistical View: Global Indicators}

The statistical view complements semantic and structural features with global indicators that are inexpensive and robust to local rewriting. We compute a vector $\mathbf{s}$ consisting of counts of risky function names (e.g., \texttt{eval}, \texttt{system}, \texttt{shell\_exec}, \texttt{base64\_decode}, \texttt{gzinflate}) matched via case-insensitive word-boundary patterns, maximal and average line length, and character-level Shannon entropy:
\begin{align}
    H(\mathbf{c}) = - \sum_{u \in \mathcal{U}} p(u)\log_2 p(u),
\end{align}
where $\mathcal{U}$ is the set of characters in $\mathbf{c}$ and $p(u)$ is the empirical frequency. We normalize $\mathbf{s}$ with batch normalization and project it with a linear layer to obtain $\mathbf{h}_{\text{stat}}$.

\subsubsection{Fusion and Training Objective}

Let $\mathbf{h}_{\text{src}}$, $\mathbf{h}_{\text{ast}}$, and $\mathbf{h}_{\text{stat}}$ denote the feature vectors from the three views. We fuse them by concatenation:
\begin{align}
    \mathbf{h} = [\mathbf{h}_{\text{src}};\mathbf{h}_{\text{ast}};\mathbf{h}_{\text{stat}}],
\end{align}
and compute logits $\mathbf{z} = f_{\psi}(\mathbf{h}) \in \mathbb{R}^2$ with an MLP classifier. The predicted probability is $p_{\psi}(y\mid \mathbf{c})=\mathrm{softmax}(\mathbf{z})_y$. The detector is trained with cross-entropy loss:
\begin{align}
    \mathcal{L}_{\text{det}}(\psi) = - \mathbb{E}_{(\mathbf{c},y)\sim \mathcal{D}} \log p_{\psi}(y\mid \mathbf{c}).
\end{align}

\subsection{Adversarial Co-Training for Generator--Detector Co-Evolution}

The central new component in ShellForge is an iterative adversarial co-training procedure that alternates between improving the generator and the detector using each other’s failures. The procedure is designed to generate increasingly evasive webshells, expand the training distribution of the detector with realistic hard cases, and prevent the detector from overfitting to explicit payload semantics by introducing safe, de-malicious variants.

\subsubsection{LLM-based De-malicious Transformation}

\begin{figure}[t]
\centering
\includegraphics[width=0.95\linewidth]{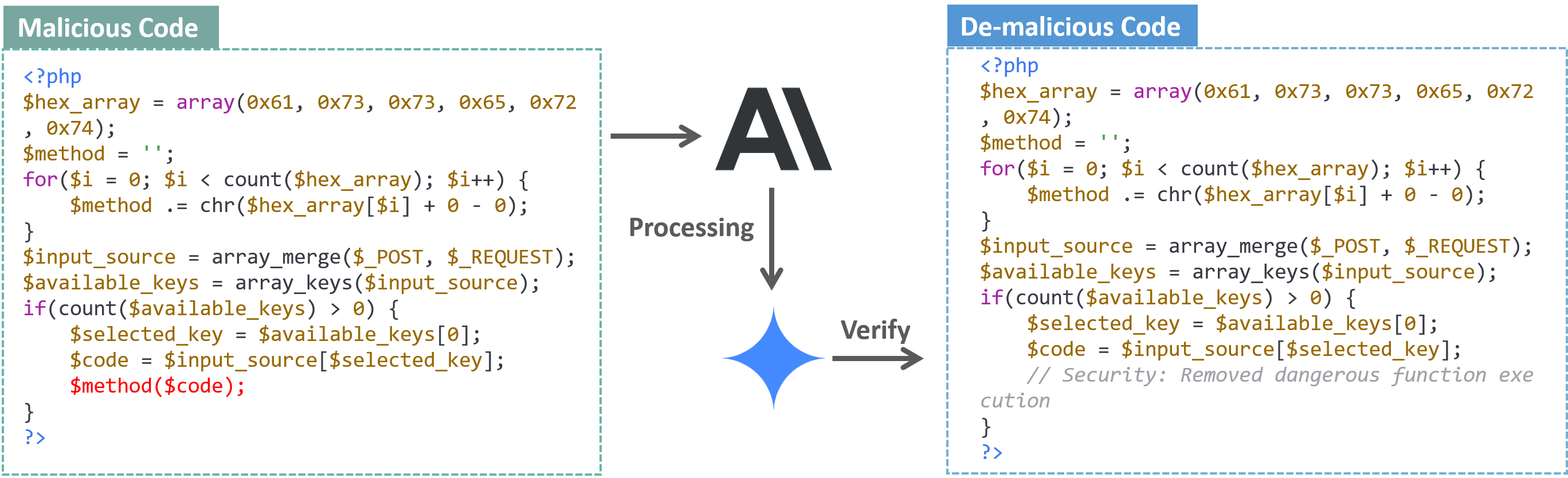}
\caption{An example of de-malicious transformation $T(\cdot)$. The red-highlighted statements that directly trigger harmful behaviors are removed, while the surrounding obfuscation patterns and code structure are preserved to obtain a benign-but-obfuscated counterpart for training.}
\label{fig:de_malicious}
\end{figure}

Given a generated webshell candidate $\tilde{\mathbf{c}} \sim G_{\phi}$, we apply a transformation $T(\cdot)$ that removes malicious behaviors while preserving obfuscation style. Concretely, $T$ rewrites security-critical behaviors (e.g., command execution paths, file write primitives, and direct evaluation) into non-malicious equivalents, while retaining the surrounding encodings, string constructions, control-flow scaffolding, and dynamic invocation patterns. This yields a benign-but-obfuscated script $\hat{\mathbf{c}} = T(\tilde{\mathbf{c}})$ that is used as a hard negative during co-training. We implement $T(\cdot)$ with an LLM-driven rewrite followed by an independent verification pass to filter out samples that still exhibit harmful behaviors. Figure~\ref{fig:de_malicious} shows an example where the harm-inducing sink is removed (highlighted in red) while the surrounding obfuscation patterns and code structure remain largely unchanged.

This design targets a key deployment risk: detectors may treat obfuscation scaffolding or the mere presence of risky APIs as a proxy for maliciousness, leading to false positives on benign-but-obfuscated code. By injecting de-malicious benign samples into training, the detector is encouraged to focus on behavior-relevant evidence rather than obfuscation artifacts.

\subsubsection{Hard-Sample Exchange Rule}

At iteration $k$, we sample a candidate set $\mathcal{G}_k$ from the current generator. For each generated candidate $\tilde{\mathbf{c}}\in\mathcal{G}_k$, the detector assigns a confidence score $q_k(\tilde{\mathbf{c}}) = p_{\psi_k}(y{=}1\mid \tilde{\mathbf{c}})$. Using a decision threshold $\delta$, we partition:
\begin{align}
    \mathcal{R}_k &= \{\tilde{\mathbf{c}}\in\mathcal{G}_k \mid q_k(\tilde{\mathbf{c}}) \ge \delta\}, \\
    \mathcal{C}_k &= \{\tilde{\mathbf{c}}\in\mathcal{G}_k \mid q_k(\tilde{\mathbf{c}}) < \delta\}.
\end{align}
Intuitively, $\mathcal{R}_k$ consists of samples that the detector can already catch (low-quality evasions), while $\mathcal{C}_k$ captures undetected evasive samples. We use these sets for bidirectional exchange: (i) for updating the generator, samples in $\mathcal{R}_k$ act as rejected signals (discouraging detectable patterns), while samples in $\mathcal{C}_k$ act as chosen signals; (ii) for updating the detector, samples in $\mathcal{C}_k$ are added as hard positives, and their de-malicious variants $T(\mathcal{C}_k)$ are added as hard negatives to reduce reliance on explicit payload semantics.

\subsubsection{Update Schedule and Stopping Criterion}

Each co-training round alternates between updating the detector by minimizing $\mathcal{L}_{\text{det}}$ on the current training set initialized from FWOID and updating the generator with a PPO-style step that incorporates the chosen/rejected signals induced by detector feedback \cite{schulman2017proximal}. We stop the adversarial procedure when both detector performance on a held-out set and generator evasiveness metrics plateau across multiple rounds, indicating diminishing returns from further co-evolution.

\section{Evaluation}
\label{sec:evaluation}

\subsection{Experimental Setup}
\paragraph{Models}
We consider two components in ShellForge: a generator $G_{\phi}$ and a detector $D_{\psi}$. For generation, we use Qwen2.5-Coder \cite{hui2024qwen2} as the base LLM and initialize it with SFT followed by PPO-style preference optimization. In this paper, we use generator-side metrics mainly to quantify the adversarial pressure imposed on the detector rather than to benchmark generation methods.
For detection, we implement a multi-view fusion model with three parallel branches: a CodeBERT encoder followed by a TextCNN module with kernel sizes 3/4/5 and 100 filters each for the source view, a Tree-sitter based AST-type sequence encoder with a 64-d embedding layer and a TextCNN module with kernel sizes 2/3/4 and 32 filters each for the AST view, and a 14-d statistical feature vector projected to 32-d through a BatchNorm+Linear layer \cite{ioffe2015batch}. The concatenated features are classified by an MLP with a 256-d hidden layer and dropout 0.3.
For the de-malicious transformation in co-evolution, we use prompt engineering with Claude Sonnet 4.5 \cite{anthropic2025claudesonnet45} to rewrite generated webshells into benign-but-obfuscated variants and use Gemini 3.0 \cite{googleai2025gemini3} as a verification pass to filter out samples that still exhibit harmful behaviors.

\paragraph{Datasets}
For the initial evaluation, we use the FWOID dataset released by \cite{wang2025incremental}, which contains 5,001 PHP webshell samples and 5,936 benign samples. Data is partitioned using a stratified 80/10/10 split into train/validation/test sets.
For adversarial co-training, we initialize the detector and generator on the FWOID training split and run iterative hard-sample exchange. During adversarial co-training, the detector and generator are initialized on the training split, and hard samples are aggregated into a co-evolution set, denoted as EvoSet. EvoSet is composed of evasive webshell samples selected as hard positives and their de-malicious benign samples $T(\cdot)$ selected as hard negatives across co-evolution rounds. Upon convergence, this collection comprises 1,138 evasive webshells and 1,090 de-malicious benign samples. When combined with FWOID, the aggregated dataset totals 6,139 webshell samples and 7,026 benign samples.

\paragraph{Metrics}
We report four metrics for detector evaluation (Accuracy, Precision, Recall, and F1), and a metric for generator evaluation (Evasion Rate).

(1) \textbf{Accuracy}: the fraction of correctly classified samples, $\mathrm{Accuracy}=\frac{N_{\text{correct}}}{N_{\text{total}}}$.

(2) \textbf{Precision}: the fraction of predicted webshells that are true webshells, $\mathrm{Precision}=\frac{\mathrm{TP}}{\mathrm{TP}+\mathrm{FP}}$.

(3) \textbf{Recall}: the fraction of webshells correctly detected, $\mathrm{Recall}=\frac{\mathrm{TP}}{\mathrm{TP}+\mathrm{FN}}$.

(4) \textbf{F1}: the harmonic mean of precision and recall, $\mathrm{F1}=\frac{2\mathrm{PR}}{\mathrm{P}+\mathrm{R}}$.

(5) \textbf{Evasion Rate (ER)}: the fraction of generated candidates not flagged by a detection engine, $\mathrm{ER}=\frac{N_{\text{undetected}}}{N_{\text{generated}}}$; we use VirusTotal \cite{virustotal} as the detection engine.

\begin{figure}[t]
\centering
\includegraphics[width=0.95\linewidth]{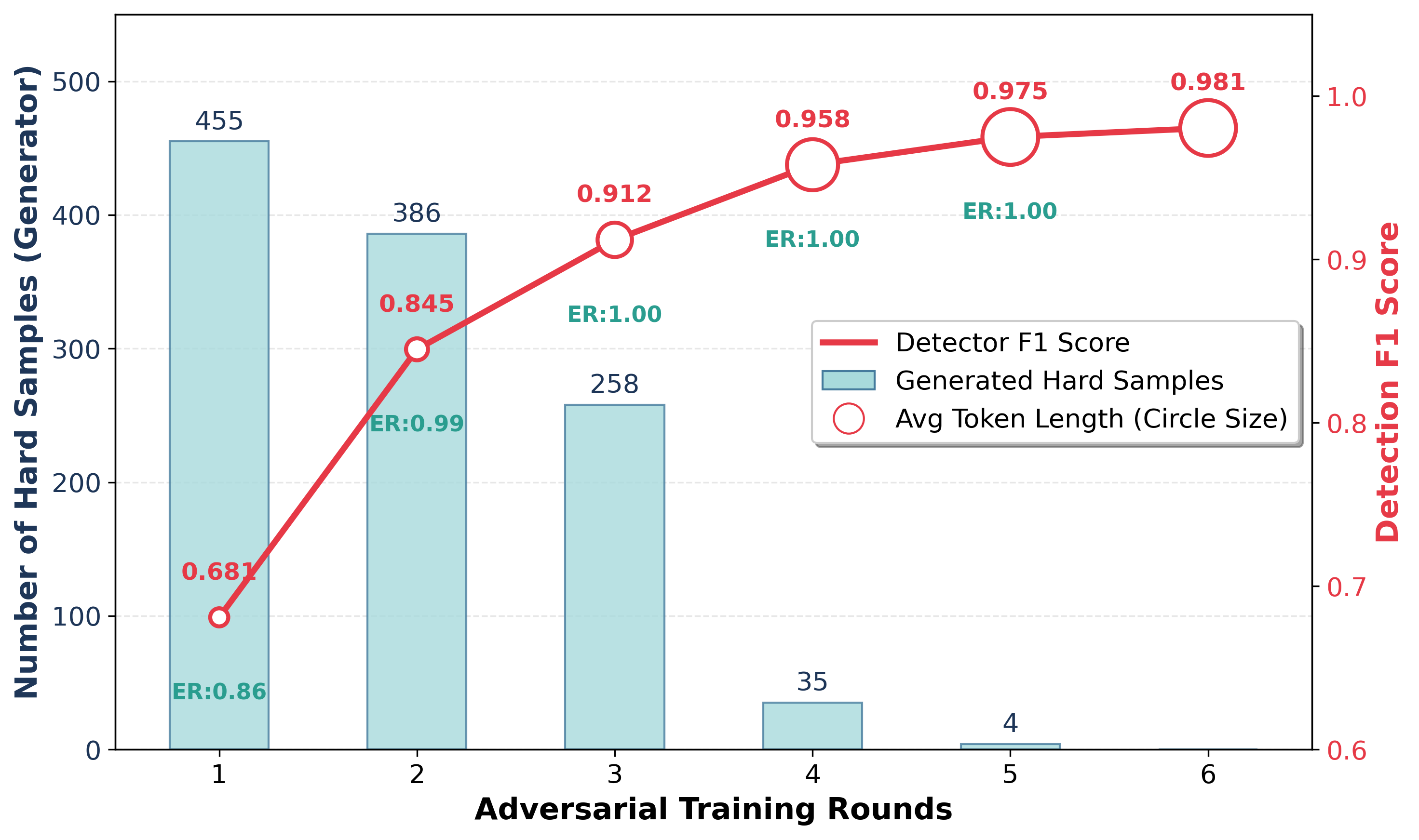}
\caption{Adversarial co-evolution over six rounds. Bars on the left axis report the number of evasive samples that bypass the current detector in each round, revealing rapid convergence as the generator exhausts obvious blind spots. The red line on the right axis tracks the detector F1 score, which increases from 0.681 to 0.981. Circle size encodes the average token length of these samples, showing that although finding samples that fool the detector becomes increasingly difficult, the remaining ones are substantially more complex. ``ER'' denotes the per-round VirusTotal evasion rate.}
\label{fig:coevolution}
\end{figure}

We report ER per round to visualize the co-evolution dynamics in Figure~\ref{fig:coevolution} and as an overall value to summarize the final generator after co-evolution. The overall ER is computed over the full set of candidates generated for evaluation after convergence.

\paragraph{Baselines}
Current webshell detection approaches typically rely on a particular program representation (e.g., lexical features, ASTs, opcodes, or program graphs) and can exhibit different robustness profiles under heavy obfuscation. We compare ShellForge’s detector with representative baselines that cover both widely used practical engines and research methods across these representations. For fairness, all baselines are evaluated on the same FWOID split and tuned on the validation set when trainable.

\begin{itemize}
\item ShellPub \cite{shellpub} is a production-grade engine combining rule matching with static code inspection. It serves as a practical baseline to evaluate how off-the-shelf detection tools handle real-world obfuscation.
\item AST-DF \cite{dong2024ast} vectorizes serialized abstract syntax tree sequences using $n$-gram and TF-IDF features. These structural representations are classified by a Deep Forest model to capture non-linear decision boundaries without neural encoders.
\item FRF-WD \cite{fang2018detecting} processes PHP opcodes via FastText to minimize sensitivity to source-level rewriting. It fuses subword embeddings with statistical indicators for classification through random forest.
\item GlareShell \cite{feng2024glareshell} constructs an interprocedural control-flow graph with risk-weighted nodes. It utilizes a Graph Neural Network with attention to capture long-range execution dependencies within obfuscated code.
\end{itemize}

\paragraph{Implementation Details}
Unless otherwise specified, all experiments are conducted on a machine equipped with an NVIDIA GeForce RTX 4070 graphics card with 8 GB of GPU memory and 32 GB of system RAM. We train the detector for 10 epochs using the AdamW optimizer \cite{loshchilov2017decoupled} with a learning rate of $2\times10^{-5}$ and a batch size of 8. For the source view, we set the maximum token length to 512 and apply head--tail truncation after semantic compression; long string literals exceeding 200 characters are replaced with \texttt{<LONG\_STR>}. For the AST view, we linearize the Tree-sitter parse tree via depth-first search (DFS) and limit the AST sequence length to 128; inline HTML is removed, string node contents are ignored, and array literals are truncated to at most 10 elements. All experiments use the same random seed for data splitting and model training. Additionally, all baseline methods are configured with the recommended hyperparameters as reported in their original papers.

\begin{figure}[t]
\centering
\includegraphics[width=0.78\linewidth]{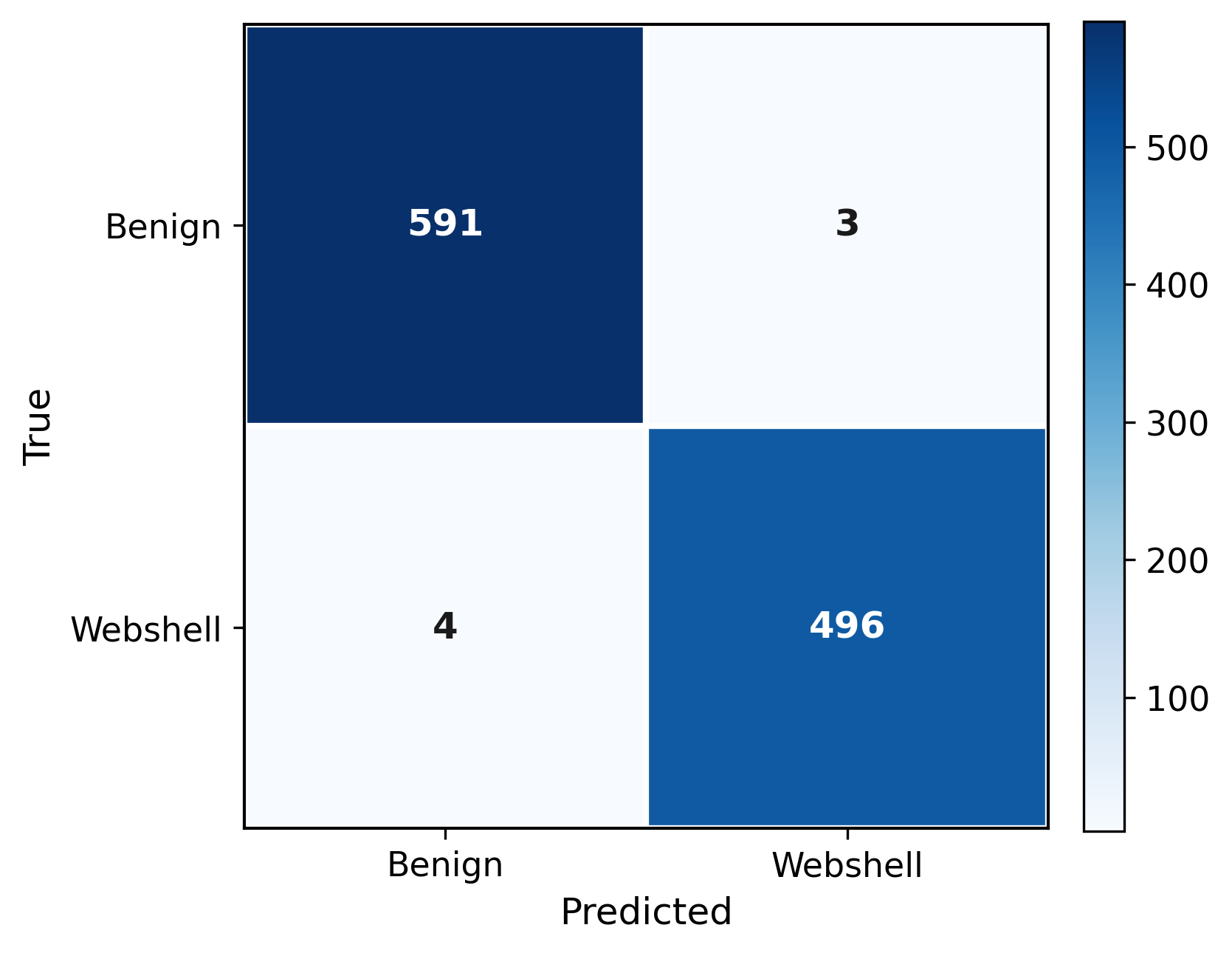}
\caption{Confusion matrix of ShellForge on the FWOID test set.}
\label{fig:confusion}
\end{figure}

\begin{figure}[t]
\centering
\includegraphics[width=0.8\linewidth]{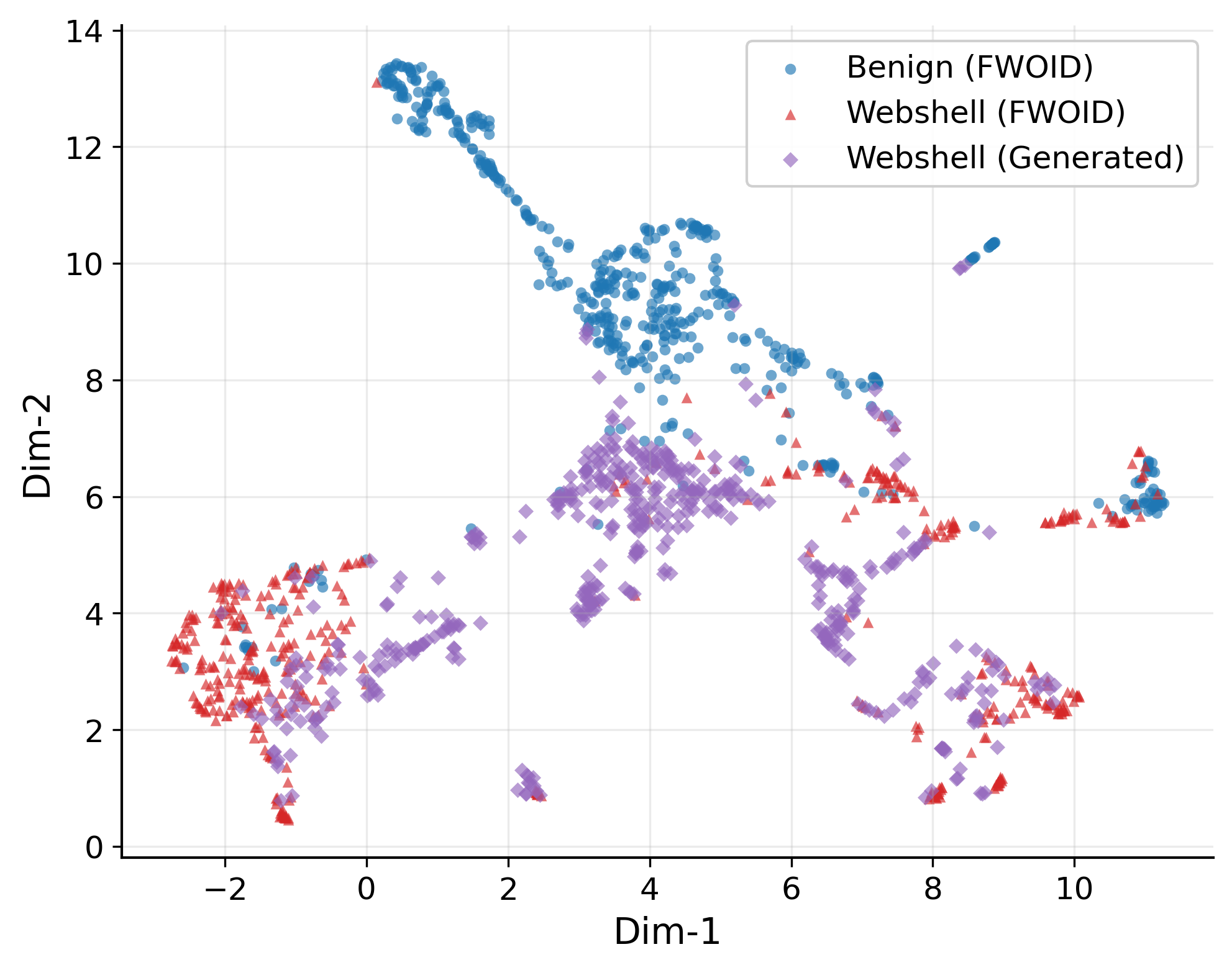}
\caption{UMAP visualization of ShellForge feature embeddings, with 500 samples each for benign (blue), original webshell (red), and generator-produced webshell (purple) groups. The generated samples populate the sparse region between benign and malicious clusters and extend along the decision boundary, indicating blind-spot filling and boundary exploration rather than simple duplication of the original webshell distribution.}
\label{fig:umap}
\end{figure}

\subsection{Results}

Table~\ref{tab:performance_fwoid} reports the detection performance on the standard FWOID test set. ShellForge achieves the best overall results across all metrics, reaching 0.9930 F1 with precision 0.9940 and recall 0.9920. Figure~\ref{fig:confusion} further confirms that the errors are rare on both classes, indicating that the multi-view fusion detector can simultaneously suppress false alarms on benign samples and avoid missing obfuscated webshells.

While strong in-distribution performance is necessary, the key question is whether a detector remains reliable when facing generator-produced evasive variants. Table~\ref{tab:performance_coevolution_gain} reports robustness on EvoSet, the hard-sample set produced during adversarial co-training. Models trained only on static FWOID suffer a severe distribution shift, with most methods dropping to around 0.55 accuracy as noted in the caption of Table~\ref{tab:performance_coevolution_gain}. After adversarial co-training initialized on FWOID and driven by hard-sample exchange, trainable baselines improve, but ShellForge attains the highest final robustness with 0.9818 accuracy and 0.9805 F1. This result supports the main claim of ShellForge: high scores under a static split can overestimate real-world robustness, whereas co-evolution continuously exposes the detector to difficult, distribution-shifting cases and reduces systematic blind spots while controlling false positives on benign-but-obfuscated code.

The co-evolution dynamics are visualized in Figure~\ref{fig:coevolution}. The number of successful evasive samples quickly decreases over rounds, suggesting that the generator progressively exhausts easy-to-find blind spots. Meanwhile, the detector’s F1 steadily increases from 0.681 to 0.981, demonstrating sustained robustness gains. Importantly, circle size grows across later rounds, indicating that although successful evasions become rarer, the remaining ones are substantially longer and more complex, forcing the detector to learn higher-order semantic patterns rather than superficial obfuscation cues.

\begin{table}[t]
\centering
\caption{Performance comparison of different detection methods on the FWOID test set. ShellForge achieves the best performance across all metrics, demonstrating the effectiveness of our proposed method.}
\label{tab:performance_fwoid}
\resizebox{0.8\textwidth}{!}{%
\begin{tabular}{lcccc}
\toprule
\textbf{Methods} & \textbf{Accuracy} & \textbf{Precision} & \textbf{Recall} & \textbf{F1 Score} \\
\midrule
ShellPub~\cite{shellpub} & 0.9073 & 0.9983 & 0.7986 & 0.8874 \\
AST-DF~\cite{dong2024ast} & 0.9845 & 0.9811 & 0.9850 & 0.9830 \\
FRF-WD~\cite{fang2018detecting} & 0.9835 & 0.9859 & 0.9780 & 0.9819 \\
GlareShell~\cite{feng2024glareshell} & 0.9872 & 0.9850 & 0.9870 & 0.9860 \\
\textbf{ShellForge (Ours)} & \textbf{0.9936} & \textbf{0.9940} & \textbf{0.9920} & \textbf{0.9930} \\
\bottomrule
\end{tabular}
}
\end{table}

\begin{table*}[t]
\centering
\caption{Comparison of detection performance on EvoSet. The main values represent the results after adversarial co-training, while the values in parentheses denote the gain $+\Delta$ compared to models trained only on the static FWOID dataset. Without co-evolution, the accuracy of most methods drops to about 0.55, indicating a severe distribution shift induced by the generated variants and de-malicious benign samples. ShellForge achieves the highest final robustness with significant performance recovery.}
\label{tab:performance_coevolution_gain}
\resizebox{\textwidth}{!}{ 
\begin{tabular}{lcccc}
\toprule
\textbf{Methods} & \textbf{Accuracy} & \textbf{Precision} & \textbf{Recall} & \textbf{F1 Score} \\
\midrule
ShellPub$^{\dagger}$~\cite{shellpub} & 0.8463 & 0.9662 & 0.6947 & 0.8083 \\
AST-DF~\cite{dong2024ast} & 0.9313 \scriptsize{(+0.4116)} & 0.9302 \scriptsize{(+0.4077)} & 0.9218 \scriptsize{(+0.2267)} & 0.9260 \scriptsize{(+0.3295)} \\
FRF-WD~\cite{fang2018detecting} & 0.9100 \scriptsize{(+0.3243)} & 0.9428 \scriptsize{(+0.3714)} & 0.8591 \scriptsize{(+0.1034)} & 0.8990 \scriptsize{(+0.2483)} \\
GlareShell~\cite{feng2024glareshell} & 0.9476 \scriptsize{(+0.4063)} & 0.9519 \scriptsize{(+0.4211)} & 0.9349 \scriptsize{(+0.0570)} & 0.9433 \scriptsize{(+0.2817)} \\
\textbf{ShellForge (Ours)} & \textbf{0.9818 \scriptsize{(+0.4297)}} & \textbf{0.9773 \scriptsize{(+0.4421)}} & \textbf{0.9837 \scriptsize{(+0.0478)}} & \textbf{0.9805 \scriptsize{(+0.2996)}} \\
\bottomrule
\multicolumn{5}{l}{\footnotesize $^{\dagger}$ShellPub is a static engine and was not retrained; thus, no gain value is reported.}
\end{tabular}
}
\end{table*}

\begin{table}[t]
\centering
\caption{Ablation study on the de-malicious strategy. We compare the model's performance on the standard FWOID test set versus a held-out de-malicious benign set that is obfuscated but non-malicious. Without the strategy, the model suffers from severe false positives, reaching 86.24\%, indicating overfitting to obfuscation patterns. The proposed strategy reduces FPR to 2.02\% with a manageable trade-off in standard F1 score.}
\label{tab:ablation_demalicious}
\resizebox{\linewidth}{!}{ 
\begin{tabular}{lcc}
\toprule
\multirow{2}{*}{\textbf{Method}} & \textbf{Standard Test (FWOID)} & \textbf{De-malicious Benign Test} \\
 & \textbf{F1 Score} & \textbf{False Positive Rate} \\
\midrule
ShellForge (w/o De-malicious) & \textbf{0.9930} & 0.8624 \\
\textbf{ShellForge (Ours)} & 0.9805 & \textbf{0.0202} \\
\bottomrule
\end{tabular}
} 
\end{table}

\begin{table}[t]
\centering
\caption{Ablation study on semantic compression and long-text truncation strategies. The combination of semantic compression and head-tail truncation (Ours) yields the best results by preserving critical logic in long webshells.}
\label{tab:ablation_longtext}
\resizebox{0.8\textwidth}{!}{
\begin{tabular}{lcccc}
\toprule
\textbf{Configuration} & \textbf{Accuracy} & \textbf{Precision} & \textbf{Recall} & \textbf{F1 Score} \\
\midrule
Head only & 0.9717 & 0.9699 & 0.9680 & 0.9690 \\
Compress only & 0.9726 & 0.9738 & 0.9660 & 0.9699 \\
Head+Tail & 0.9726 & 0.9719 & 0.9680 & 0.9699 \\
Head+Compress & 0.9890 & 0.9939 & 0.9820 & 0.9879 \\
\textbf{ShellForge (Ours)} & \textbf{0.9936} & \textbf{0.9940} & \textbf{0.9920} & \textbf{0.9930} \\
\bottomrule
\end{tabular}
}
\end{table}

\begin{table}[t]
\centering
\caption{Ablation study on the contribution of different views. While single views have limitations (e.g., Stats View), fusing Source, AST, and Statistical features significantly improves detection performance.}
\label{tab:ablation_multiview}
\resizebox{0.8\textwidth}{!}{
\begin{tabular}{lcccc}
\toprule
\textbf{Feature Combinations} & \textbf{Accuracy} & \textbf{Precision} & \textbf{Recall} & \textbf{F1 Score} \\
\midrule
Source View only & 0.9735 & 0.9758 & 0.9660 & 0.9709 \\
AST View only & 0.8684 & 0.7957 & 0.9580 & 0.8693 \\
Stats View only & 0.7578 & 0.9434 & 0.5000 & 0.6536 \\
Source + AST & 0.9790 & 0.9857 & 0.9680 & 0.9768 \\
Source + Stats & 0.9835 & 0.9879 & 0.9760 & 0.9819 \\
\textbf{ShellForge (Ours)} & \textbf{0.9936} & \textbf{0.9940} & \textbf{0.9920} & \textbf{0.9930} \\
\bottomrule
\end{tabular}
}
\end{table}

To further illustrate why co-evolution improves robustness, Figure~\ref{fig:umap} visualizes the feature embeddings of benign, original webshell, and generator-produced samples. The generated samples concentrate in the sparse region between the benign cluster and the original webshell cluster, and extend along the boundary region. This blind-spot filling and boundary exploration pattern indicates that the generator does not simply duplicate existing webshells, but discovers previously underrepresented hard cases near the decision boundary, which provides informative training signals for hardening the detector.

\subsection{Ablation Study}

This section examines the contribution of key design choices in ShellForge, focusing on the de-malicious strategy used during co-evolution, long-text handling in the source view, and the effect of multi-view fusion.

Table~\ref{tab:ablation_demalicious} evaluates the de-malicious strategy using both the standard FWOID test set and a held-out de-malicious benign set of 1,090 samples that is highly obfuscated but non-malicious. This setting captures a common deployment case: benign PHP code may contain risky APIs or resemble obfuscated templates, and a detector that equates obfuscation with maliciousness can raise false alarms even when no harmful behavior is present. Without de-malicious negatives, the detector attains 0.9930 F1 on FWOID but produces an extremely high false positive rate of 86.24\% on the de-malicious benign set, indicating over-reliance on obfuscation artifacts. With the proposed strategy, the standard-set F1 drops to 0.9805, yet the false positive rate collapses to 2.02\%. We consider this trade-off worthwhile because it converts a brittle detector that over-alerts on benign-but-obfuscated code into a robust detector that better reflects real-world operating conditions.

Table~\ref{tab:ablation_longtext} studies semantic compression and truncation strategies for long webshell files. Naive head-only truncation yields clear degradation, because critical payload logic can appear near the end of a script. Semantic compression alone or head+tail truncation alone provides limited gains, while combining compression with head--tail retention improves performance to 0.9879 F1. The full ShellForge pipeline achieves the best result, reaching 0.9930 F1, supporting the claim that preserving both key logic segments and removing excessive string noise is important for robust long-context webshell detection.

Table~\ref{tab:ablation_multiview} quantifies the effect of fusing multiple views. Single-view models exhibit complementary weaknesses: the AST-only branch has high recall but poor precision under obfuscation, while the statistics-only branch performs substantially worse due to its limited expressiveness. Two-view combinations improve stability, but the best performance is achieved when all three views are fused, reaching 0.9930 F1. This confirms that the semantic, structural, and statistical views provide non-redundant signals and jointly contribute to ShellForge’s robustness.

\section{Conclusion}

In this paper, we propose ShellForge, an adversarial co-evolution framework for robust PHP webshell defense. ShellForge couples a trainable webshell generator with a multi-view fusion detector, and drives their mutual improvement through iterative hard-sample exchange under a safety-preserving de-malicious transformation. By integrating semantic, structural, and statistical views and continuously expanding training coverage around the decision boundary, ShellForge strengthens detection robustness against rapidly evolving, heavily obfuscated webshell variants. Extensive experiments on a public benchmark and an adversarially expanded setting validate the effectiveness of our framework, showing consistently strong performance and improved robustness compared to existing methods.


\appendix


\bibliographystyle{elsarticle-num} 
\bibliography{ref}





\end{document}